\documentclass[11pt]{article}

\usepackage{amsmath,amssymb}
\usepackage{graphicx}

\newcommand{\be}{\begin{equation}}
\newcommand{\ee}{\end{equation}}
\newcommand{\bea}{\begin{eqnarray}}
\newcommand{\eea}{\end{eqnarray}}
\newcommand{\ba}{\begin{array}}
\newcommand{\ea}{\end{array}}

\def \nn {\nonumber}
\newcommand{\eq}[1]{(\ref{#1})}

\newcommand{\alp}{\alpha'}

\newcommand{\p}{\partial}

\newcommand{\cM}{{\cal{M}}}

\newcommand{\cO}{{\cal{O}}}

\newcommand{\cZ}{{\cal{Z}}}

\begin{document}

\begin{titlepage}
\begin{flushright}
\normalsize  June 1, 2007
\end{flushright}

\vfill

\begin{center}
{\large \bf
Hagedorn Strings and Correspondence Principle in AdS$_{3}$}
\vfill
{
Feng-Li Lin\footnote{\tt linfengli@phy.ntnu.edu.tw}
},
{
Toshihiro Matsuo\footnote{\tt tmatsuo@home.phy.ntnu.edu.tw}
} and
{
Dan Tomino\footnote{\tt dan@home.phy.ntnu.edu.tw}
}

\bigskip
{\it
Department of Physics,
National Taiwan Normal University, \\
Taipei City, 116, Taiwan
}

\end{center}

\vfill
\begin{abstract}
Motivated by the possibility of formulating a strings/black hole correspondence in AdS space, 
we extract the Hagedorn behavior of thermal AdS$_3$ bosonic string from 1-loop partition function of $SL(2,R)$ WZW model. 
We find that the Hagedorn temperature is monotonically increasing as the AdS radius shrinks, 
reaches a maximum of order of string scale set by the unitarity bound of the CFT for internal space. 
The resulting density of states near the Hagedorn temperature resembles the form as  for strings in flat space and is dominated by the space-like long string configurations. 
We then argue a conjectured strings/black hole correspondence in AdS space by applying the Hagedorn thermodynamics. 
We find the size of the corresponding black hole is a function of the AdS radius. 
For large AdS radius a black hole far bigger than the string scale will form.
On the contrary, when the AdS and string scales are comparable a string size black hole will form. 
We also examine strings on BTZ background obtained through $SL(2,Z)$ transformation. 
We find a tachyonic divergence for a BTZ black hole of string scale size.

\end{abstract}
\vfill

\end{titlepage}

\setcounter{footnote}{0}

\section{Introduction}

The critical behavior of string theory near the Hagedorn temperature is known to be relevant to the strings/black hole correspondence \cite{Susskind:1993ws}, \cite{Horowitz:1996nw}.
Especially, the Hagedorn entropy roughly matches the Bekenstein-Hawking one at the correspondence point, i.e. when the Hagedorn and Hawking temperatures are of the
same order.
Moreover, the Hagedorn string can be treated as a thermal scalar \cite{Brandenberger:1988aj} due to the appearance of the Euclidean time-like unit-winding tachyon \cite{Sathiapalan:1986db}, \cite{Kogan:1987jd}, \cite{Atick:1988si}.
The above results were done in the flat space.
It is interesting to understand the Hagedorn behavior in curved space where some intrinsic curvature scale may play a role in understanding the $\alpha'$ effect.
Among these, the Anti-de Sitter (AdS) Hagedorn string is the most interesting in light
of the AdS/CFT correspondence.

At low energy, there is a Hawking-Page transition occurring at temperature $T_{HP}\sim 1/l_{AdS}$ such that the thermal AdS space condenses to AdS-Schwarzschild black hole.
The Hawking-Page temperature $T_{HP}$ is far below the Hagedorn temperature $\beta_s^{-1}$.
However, if we superheat the AdS strings above $T_{HP}$ and approach $\beta_s^{-1}$, we may wonder how the effect of the curvature scale $k=(l_{AdS}/l_s)^2$ comes into play and affects the Hagedorn behavior.
Naively, the AdS curvature scale provides a finite size effect of the ambient space, so that one would like to see if the strings/black hole correspondence principle in flat space can be generalized to the AdS case.
To give some clue in answering this question, we will compare the Hagedorn entropy and the Bekenstein-Hawking one in AdS space.
We re-write the Bekenstein-Hawking entropy of the AdS$_{d+1}$-charged Schwarzschild black hole $(d>2)$ in the following suggestive form
\bea\label{SBH}
S_{BH}=
\frac{d-2}{d-1}\frac{1+\frac{d}{d-2}\frac{r_+^2}{l_{AdS}^2}-\mu^2}{1+\frac{r_+^2}{l_{AdS}^2}+\mu^2}
\beta_{BH}M ,
\eea
where $M$ and $r_+$ are the mass and size of horizon of the black hole.
The inverse Hawking temperature $\beta_{BH}$ and the chemical potential $\mu$ conjugate to the charge $Q$ are written respectively as
\be
\beta_{BH}=\frac{4\pi l_{AdS}^2 r_+}{(d-2)(1-\mu^2)l_{AdS}^2+dr_+^2}, \qquad
\mu=\frac{w_{d+1} Q}{2r_+^{d-2}}
\ee
where $w_{d+1}$ is a constant related to the Newton constant in $d+1$ dimensions.
Note that the pre-factor in front of $\beta_{BH} M$ in \eq{SBH} is always of order one for any value of $r_+/l_{AdS}$, that includes the charged Schwarzschild for $r_+/l_{AdS}\sim 0$.
This suggests that the underlying density of states takes the form as $\rho \sim e^{\beta(M) M}$, not as particle-like, i.e. $\rho\sim M^{\alpha}$ that yields $S \sim \ln M$.

Compare \eq{SBH} with the expected Hagedorn entropy of string
\be
S_{string}=\beta_s(k)M,
\ee
we may have strings/black hole correspondence whenever $\beta_s\sim
\beta_{BH}$ so that $S_{BH}\sim S_{string}$. This then generalizes the strings/black hole correspondence in the flat space to AdS space (see section \ref{correspondence} for more arguments).

Similarly, for $d=2$, the entropy of the BTZ black hole is
\be
S_{BH}=2\frac{1-(l_{AdS}\mu)^2}{1+(l_{AdS}\mu)^2}\beta_{BH}M
\ee
where the Hawking temperature and  the chemical potential conjugate to the angular momentum $J$ are
\bea
\beta_{BH}=\frac{2\pi l_{AdS}^2}{r_+}\frac{1}{1-(l_{AdS}\mu)^2} , \quad
\mu=4G_NJ/r_+^2 ,
\eea
respectively, where $G_N$ is the Newton constant.
This is also Hagedorn-like.

Motivated by the strings/black hole correspondence, it is interesting to explore the Hagedorn behavior of AdS strings and extract the dependence of Hagedorn temperature and the density of states on the AdS curvature scale.
However, this is hard because the AdS string theory usually has not been exactly solved. Despite that, in \cite{Barbon:2004dd,Kruczenski:2005pj} the authors discuss the issue for AdS$_5$ case by approximate methods.
They argue that the Euclidean AdS black hole can be thought as the condensation of the thermal scalar in a compact space, which then have the Hagedorn-like density of states.
More recently, a nice paper \cite{LiuH1} in developing the techniques of re-summing the worldsheet diagrams of the AdS string in the Hagedorn regime has been done by introducing the double scaling limit, and the relation between Hagedorn behavior of AdS string and its dual Super-Yang-Mills (SYM) theory is explicitly uncovered.
Earlier discussions on the Hagedorn strings by exploiting the Hagedorn behavior of the dual SYM theory \cite{sundborg} can be found in \cite{dsym1, dsym2, dsym3}.

 Instead, in this work we will discuss the Hagedorn behavior of the bosonic string theory on Euclidean $AdS_3 \times \cM$, which is exactly solvable \cite{Maldacena:2000hw} and whose 1-loop partition function was given in \cite{Maldacena:2000kv}.
By directly studying the partition function, we can see how the curvature scale $k$ affects the Hagedorn behavior, and extract Hagedorn temperature and the density of states. 
Our main result is following. 
The Hagedorn inverse temperature and the leading behavior of the density of states of the Hagedorn string in the micro-canonical ensemble are
\bea
\beta_s=\beta_{AdS}:= \sqrt{\frac{2\pi^2 \alpha' (c_{int}+1)}{3}}=4\pi  l_s\sqrt{\frac{k-9/4}{k-2}}
\eea
and
\be
\Omega(E)\propto {e^{\beta_{AdS}E} \over E^{(c_{int}+1)/2+1}} ,
\ee
respectively.
In the above we have assumed $c_{int}$-dimensional ``internal space" $\cM$ is flat and non-compact.
An essential difference from the flat space string is the appearance of infinite new long string degrees of freedom \cite{Maldacena:2000hw,Maldacena:2000kv},
and it is interesting to see how the Hagedorn thermodynamics encodes these topological information as the case in flat space. 
Moreover, for unitary internal CFT the value of $k$ should be larger than $2+\frac{6}{23}$ $(>\frac{9}{4})$. 
This implies that the Hagedorn temperature cannot be infinite.

We will organize our paper as follows.
In section 2 we will briefly review the path integral formulation of thermal AdS$_3$ string theory, and set up our notations.
In section 3, we first discuss the pole structure of the integrand of the partition function in subsection 3.1.
Due to the existence of the long string configuration spreading over the AdS space, the pole structure is more complicated than the one for the flat space string.
In subsection 3.2 we extract the Hagedorn temperature and the corresponding density of states.
We obtain the explicit $k$-dependence of the Hagedorn temperature, which is monotonically increasing as the AdS curvature grows.
Moreover, we can identify the contribution of the long string to the Hagedorn spectrum.
Based on the partition function in the Hagedorn regime obtained in section 3, we discuss corresponding Hagedorn thermodynamics in section 4, which is in close resemblance to the case in flat space.
In section 5 we discuss some implication of our results to a conjectured strings/black hole correspondence in AdS$_3$ space.
In \cite{Giveon:2005mi} some aspects on the correspondence between strings in AdS$_3$ and BTZ black hole are uncovered by taking a different approach in that the parameter $k$ is varied.
Instead here we examine the theory by changing temperature $\beta^{-1}$ with fixed $k$.
In section 6, we investigate the Hagedorn behavior of the strings in BTZ black hole by $SL(2, Z)$ transformation acting on the boundary torus of AdS$_3$.
In section 7 we briefly comment on the case with nonzero chemical potential and conclude our work. 
In Appendix we give a technical discussion about the sub-dominant contribution to the Hagedorn partition function.

   {\bf Noted added in proof:} In the final stage of drafting this paper, we are informed a related on-going work \cite{Berkooz:2007fe} \footnote{See also \cite{Rangamani:2007fz}.}.   

\section{AdS$_3$ string and its thermal partition function}

The bosonic string theory in AdS$_3$ space can be realized as a $SL(2,R)$ WZW model and is solvable.
The spectra and 1-loop thermal partition function have been studied in
\cite{Maldacena:2000hw} and \cite{Maldacena:2000kv} respectively.
In these papers, it was pointed out that the complete string spectra are generated by the
spectral flow symmetry of the theory.
Moreover, the continuous branches of the spectra represent the long string states extending along the radial direction to the AdS boundary.
These long string spectra-flowed states appear as poles in the partition function,
which are absent in the flat space case
\footnote{Strictly speaking, the poles exist at the boundary of moduli in the flat space case.}.
These new poles will play an essential role in the Hagedorn regime as will be shown.

We consider the string on $AdS_3 \times {\cal M}$, where an internal
manifold ${\cal M}$ is described by a sigma model that provides appropriate
central charge. The worldsheet conformal field theory contains
three part, the one for AdS$_3$, the one for ${\cal M}$ and the $(b,c)$
ghosts. The 1-loop partition function has been studied in
\cite{Maldacena:2000kv}.
Here we will review it and set up the convention of the notations.
We first consider the AdS$_3$ part and then combine the rest twos.

The metric on Euclidian AdS$_3$ in the global coordinates $(\rho, t,
\theta)$, all of which are dimensionless, is
\bea
\frac{ds^2}{\alp k}=\cosh^2\rho dt^2+d\rho^2+\sinh^2\rho d\theta^2 ,
\eea
where $k=l_{AdS}^2/\alp$ is a dimensionless number.
It is convenient to use coordinates $(V, \bar{V}, \Phi)$ defined as
\bea\label{coor-t}
V=\sqrt{\alp k}\sinh \rho e^{i\theta}, \quad
\bar{V}=\sqrt{\alp k}\sinh \rho e^{-i\theta}, \quad
\Phi=\sqrt{\alp k}(t-\ln \cosh \rho),
\eea
in which the metric becomes
\bea
ds^2=d\Phi^2
+\left(dV+\frac{Vd\Phi}{\sqrt{k}}\right)\left(d\bar{V}+\frac{\bar{V}d\Phi}{\sqrt{k}}\right).
\eea
This becomes the flat metric in the limit $k \to \infty$\footnote{In
contrast, the metric in \cite{Maldacena:2000kv} does not reduce to
flat space one in the limit $k \to \infty$. Their coordinates are
related to ours in \eq{coor-t} by an overall scaling by
$\sqrt{k}$.}. Thermal AdS$_3$ is defined by the following
identifications in the global coordinates
\bea
t+i \theta \sim t + i\theta+\beta/\sqrt{\alp k}+i\mu \beta ,
\eea
were $\beta$ is the inverse temperature of thermal AdS$_3$ and $i\mu$ is the
imaginary chemical potential for the angular momentum.
 We impose the identifications
\bea
V \sim V e^{i\mu\beta}, \quad
\bar{V} \sim \bar{V}e^{-i\mu \beta},
\quad
\Phi \sim \Phi+\beta.
\label{sym1}
\eea

The WZW action in these coordinates is
\bea
S = \frac{1}{\pi\alp}\int d^2z
\left[\p \Phi \bar{\p} \Phi
+\left(\p \bar{V}+\frac{\p \Phi \bar{V}}{\sqrt{k}}\right)\left(\bar{\p}V +\frac{\bar{\p}\Phi V}{\sqrt{k}}\right)
\right] .
\eea
Due to the identification \eq{sym1}, the boundary conditions on the worldsheet torus become
\bea
\Phi(z+2\pi)=\Phi(z)+\beta n,&& \quad \Phi(z+2\pi\tau)=\Phi(z)+\beta m,
\nonumber \\
V(z+2\pi)=V(z)e^{in\mu \beta},&& \quad V(z+2\pi\tau)=V(z)e^{im\mu \beta} ,
\eea
where $n,m \in Z$.
As shown in \cite{Maldacena:2000kv} by adopting the technique in dealing with path integral of $H_3$ WZW model \cite{Gawedzki:1991yu}, one can exactly evaluate the partition function of the AdS sector, and the result is
\bea\label{AdS3Z}
\cZ_{AdS}(\beta, \mu ;\tau)
=\frac{\beta(1-2/k)^{1/2}}{(4\pi^2\alp\tau_2)^{1/2}} \sum_{(n,m)\ne
(0,0)}
\frac{e^{-\beta^2|m-n\tau|^2/4\pi\alp\tau_2+2\pi({\mbox{Im}}U_{n,m})^2/\tau_2}}{|\vartheta_1(\tau,U_{n,m})|^2},
\eea
where
\bea
U_{n,m}(\tau)
=\frac{i\beta(n \bar{\tau}-m)}{2\pi\sqrt{\alp}} \left(\frac{1}{\sqrt{k}}-i \mu\sqrt{\alp}\right)  ,
\eea
and the theta function is defined as
\bea\label{theta11}
\vartheta_1(\tau,U)=2q^{1/8} \sin(\pi U) \prod_{m=1}^\infty(1-q^m)(1-zq^m)(1-z^{-1}q^m) ,
\eea
where $q=e^{2\pi i \tau}, z=e^{2\pi iU}$.
Note that the $(m,n)=(0,0)$-sector is modular invariant and corresponding to zero temperature contribution, therefore we can neglect it in our discussion about Hagedorn behavior.
Moreover, in \eq{AdS3Z} there is a shift $2/k$ in the overall factor $(1-2/k)^{1/2}$ due to the chiral anomaly arising in the integration of $V$ and $\bar{V}$.
To ensure reality of $Z_{AdS}$, we should require $k\ge 2$.
Here we also like to point out  that only the temperature direction has the momentum zero-mode whose (Gaussian) integration giving rise to the factor $(4\pi^2 \alp \tau_2)^{-1/2}$ in (\ref{AdS3Z}).
The radial and angular directions do not have such kind of zero-modes, thus provide no corresponding factor.

The ghost part contribution to the partition function is
\bea
\cZ_{gh}(\tau)=|\eta(\tau)|^4=(q\bar{q})^{2/24}|\prod_{n=1}^\infty(1-q^n)|^4 .
\eea

We also have the contribution from the internal space which will depend on the type of internal space.
Moreover, there is constraint on the central charge $c_{int}$ of internal CFT, that is,  the central charge $c_{int}$ should satisfy the anomaly free condition
\be
c_{SL(2,R)}+c_{int}=26.
\ee
Since the central charge for the AdS$_3$ sector is $c_{SL(2,R)}=3+\frac{6}{k-2}$, thus the central charge of the internal CFT should be
\be
c_{int}=23-\frac{6}{k-2}.
\ee
For the internal CFT to be unitary, we should require $c_{int}\ge 0$, that is, we should constraint $k$ by a lower bound as following
\be\label{k0e}
k\ge k_0=2+{6\over 23}.
\ee

For simplicity and concreteness, we assume that the internal space ${\cal M}$ is flat and non-compact and its corresponding CFT is described by $c_{int}$ free bosons \footnote{One can consider more general types of CFTs for internal space, however, the detailed Hagedorn thermodynamics will depend on the choice. We just choose the free boson for explicit calculations in the later sections.}.
Therefore, the corresponding partition function is
\bea\label{zint0}
\cZ_{int}(\tau)=(q\bar{q})^{-c_{int}/24}\sum_{h,\bar{h}}D(h,
\bar{h})q^h \bar{q}^{\bar{h}}=V_{int}(4\pi^2 \alp
\tau_2)^{-c_{int}/2}|\eta(\tau)|^{-2c_{int}}
\eea
where $V_{int}$ is the volume of the internal space.
Here we also assume there is no non-trivial cycle in ${\cal M}$ around which string can wind.

The total partition function is obtained by integrating the product of the above partition functions over the fundamental region, i.e., $F_0$, that is
\bea
Z(\beta) &=& \int_{F_0} \frac{d^2\tau}{4\tau_2}\cZ_{gh} \cZ_{int} \cZ_{AdS}
\nonumber \\
&=&V_{int}\int_{F_0} \frac{d^2\tau}{4\tau_2}
\frac{\beta(1-2/k)^{1/2}|\eta(\tau)|^{4-2c_{int}}}{(4\pi^2\alp\tau_2)^{(c_{int}+1)/2}}
\sum_{n,m=-\infty}^{\infty}
\frac{e^{-\beta^2|m-n\tau|^2/4\pi\alp\tau_2+2\pi({\mbox{Im}}U_{n,m})^2/\tau_2}}{|\vartheta_1(\tau,U_{n,m})|^2} ,
\eea
where $(n,m)\ne (0,0)$ in the sum is understood.

As shown in \cite{Polchinski:1985zf,McClain:1986id}, using the fact that $(n,m)$ transforms as $SL(2,Z)$ doublet, we employ the $SL(2,Z)$ transformation to map the fundamental domain into the strip $R: \tau_2> 0, |\tau_1|\le 1/2$, so that we can keep only $n=0$ term in the partition function, and change the modular integration over $F_0$ to the one over $R$.

Moreover, in dilute gas approximation it is enough to consider only the single string partition function which is given by $m=\pm 1$:
\bea\label{Z1}
Z_1(\beta) = 2V_{int}\beta(1-2/k)^{1/2}\int_{R}
\frac{d^2\tau}{4\tau_2}
\frac{|\eta(\tau)|^{4-2c_{int}}}{(4\pi^2\alp\tau_2)^{(c_{int}+1)/2}}
\frac{e^{-(1-2/k)\beta^2 /4\pi\alp\tau_2}}
{|\vartheta_1(\tau,U_{0,1})|^2} ,
\eea
and the partition function of the string gas as follows (see e.g. \cite{Kruczenski:2005pj})
\be\label{multistringp}
Z(\beta)\simeq e^{Z_1(\beta)}.
\ee

In the next section we will extract the Hagedorn behavior from
this single string partition function.

\section{AdS$_3$ strings at Hagedorn temperature}
In this section we would like to show the existence of the Hagedorn phase in AdS$_3$ space by the saddle-point approximation, and then extract the density of states out of the partition function, which is relevant in determining the Hagedorn thermodynamics.
As for the flat space string, these information are encoded in the UV/IR regime in the integrand of the partition function over the strip domain $R$.
However, as we will see that we have additional poles in the moduli associated with the long strings than in the flat space, we need to check if these poles make contribution to the Hagedorn density of states or not.

In the following we will set $\alp =l_s^2=1$ for simplicity.
This is equivalent to measure the inverse temperature $\beta$ by the unit string length $l_s$.
With this, recall that
\be
U_{0,1}=-{\mu\beta \over 2\pi}-{i\beta \over 2\pi \sqrt{k}}=:U.
\ee

\subsection{The location of poles}

From the expression of the single string partition function \eq{Z1}, the poles of the integrand are encoded in the zeros of $\vartheta_1(\tau,U)$.
Especially, we are interested in the Hagedorn regime, which turns out to be in the limit of $\tau_1, \tau_2 \to 0$. We will study the behavior of $\vartheta_1(\tau,U)$ and extract the Hagedorn behavior.
To this end, it is convenient to use a form after modular transformation.
The modular property of theta function
\footnote{
Modular transformation is
\bea
\vartheta_1(\tau,U)=i(-i\tau)^{-1/2}\exp(-\pi i
U^2/\tau)\vartheta_1(-1/\tau, U/\tau) .
\eea}
yields
\bea\label{theta12}
|\vartheta_1(\tau, U)| =
2|\tau|^{-1/2}\Big|
e^{-\frac{\pi i}{4\tau}}
e^{-\frac{\pi iU^2}{\tau}}
\sin\left(\frac{\pi U}{\tau}\right)
\prod_{r=1}^\infty
(1-e^{-\frac{2\pi i r}{\tau}})
(1-e^{-\frac{2\pi i(r+U)}{\tau}})
(1-e^{-\frac{2\pi i(r-U)}{\tau}})
\Big|.
\eea
 We see that the theta function has infinite number of zeros which give rise to poles of the partition function. We will investigate the locations of these poles over the strip domain.
 Without loss of generality we require $\mu \geq 0$.

  First of all, let us examine the poles coming from the zeros in the factors
$\prod_{r=1}^\infty(1-e^{-2\pi i r/\tau})$ (which is common in $\cZ_{int}$ and $\cZ_{gh}$).
These poles also appear in the flat space string.
They locate on the $\tau_1$-axis as follows
\be\label{flat pole}
\tau_1={r\over w}, \qquad \tau_2=0, \quad w= \pm 1, \pm 2,\cdots , \quad r= 1, 2,\cdots .
\ee
Here we should only include the poles inside the strip domain, i.e. $\tau_2> 0, |\tau_1|\le 1/2$.
As will be shown later, the integrand for the partition function has the saddle-point located at $\tau_1=0$,
which does not appear in \eq{flat pole}.
Therefore, these poles do not play essential role at the Hagedorn regime.

Now we will consider the poles which are absent in the flat space string but arise here due to the compactness of the AdS space. As shown in \cite{Maldacena:2000hw,Maldacena:2000kv} these additional poles are associated with the long strings which transverse the AdS space. Let us examine the zeros appear in the factors
$\prod_{r=1}^\infty(1-e^{-2\pi i(r+U)/\tau})$.
The exponents are
\be\label{zeros1}
2\pi i(r+U)=\frac{1}{|\tau|^2}\left[
-\left(2\pi r-\mu\beta\right)\tau_2
-\frac{\beta}{\sqrt{k}}\tau_1
-i\left\{
\left(2\pi r - \mu\beta\right)\tau_1
-\frac{\beta}{\sqrt{k}}\tau_2
\right\}
\right] .
\ee
To yield a zero in the factors, the real part has to vanish, i.e.,
\bea\label{realpart}
{\tau_2 \over \tau_1}
=-\frac{\beta}{(2\pi r- \mu\beta)\sqrt{k}} ,\qquad r=1,2,\cdots
\eea
which define ``pole lines" of slope $\tau_2/\tau_1$ labeled by $r$.
The locations of poles on each pole line are determined by the condition that the imaginary part of \eq{zeros1} should be $2\pi w, w \in Z$.
Together with the condition \eq{realpart}, it is easy to see that the poles from the factor $\prod_{r=1}^\infty(1-e^{-2\pi i(r+U)/\tau})$ are located at
\bea\label{factor1}
\tau_1=-\frac{1}{2\pi w}(2\pi r-\mu\beta), \qquad
\tau_2=\frac{1}{2\pi w}\frac{\beta}{\sqrt{k}}, \quad w=0, 1, 2,\cdots , \quad r=1, 2, \cdots .
\eea
Again we should only include the poles inside the strip domain. For example, if $\mu=0$, the $w=0$ pole should be ruled out.

Similarly, we can read off the pole lines and the poles on them from the factors in $\prod_{r=1}^\infty(1-e^{-2\pi i(r-U)/\tau})$. The pole lines have the slope
\bea
{\tau_2 \over \tau_1}
=\frac{\beta}{(2\pi r+\mu\beta)\sqrt{k}}, \qquad r=1,2,\cdots
\eea
and the poles locate at
\bea
\tau_1=\frac{1}{2\pi w}(2\pi r+\mu\beta), \qquad
\tau_2=\frac{1}{2\pi w}\frac{\beta}{\sqrt{k}}, \quad w=0, 1, 2,\cdots , \quad r=1,2,\cdots .
\eea
\begin{figure}[h]
\begin{center}
\resizebox{8cm}{8cm}{\includegraphics{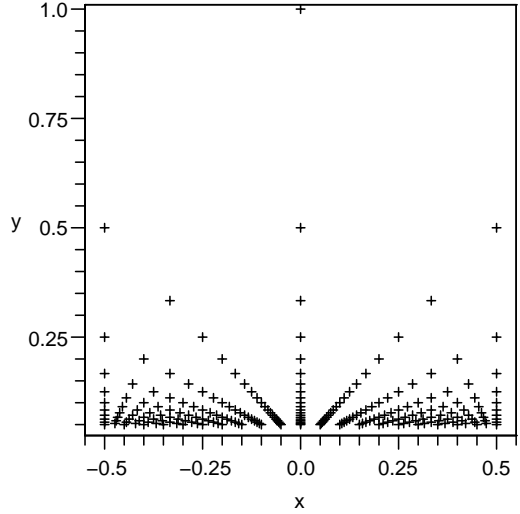}}
\end{center}
\caption{The pole structure of the integrand of the partition function. The strip domain $R$ is divided into many pairs of the nearly-triangular-strip sub-domains labeled by $\Delta_{s}$.}
\label{fig:pole}
\end{figure}
For $\mu=0$ the $w=0$ pole is outside the strip domain.

 Finally, the sin factor has a pole line with slope
\be
{\tau_2\over \tau_1}={1\over \mu \sqrt{k}}
\ee
on which the poles are distributed as
\be
\tau_1=\frac{\mu\beta}{2\pi w}, \qquad
\tau_2=\frac{1}{2\pi w}\frac{\beta}{\sqrt{k}}, \quad w=0, 1, 2,\cdots.
\ee
For $\mu=0$, the poles are understood to be located on the positive $\tau_2$-axis, and $w=0$ pole is located at $\tau_1=0,\tau_2=\infty$.

It is interesting to note that the pattern of the pole lines is invariant under $r\to r\pm 1$, this implies that it is enough to consider the range $0 \leq \mu\beta < 2\pi$. This is consistent with our earlier requirement in \eq{sym1} that $\mu \beta$ is an angular parameter.

\subsection{Evaluating the partition function}
It will turn out that the information about the high temperature behavior of string is encoded in the small $\tau_2$ region in the moduli integral (\ref{Z1}).
The asymptotic behavior of its integrand indeed depends on how it approaches the origin.
We decompose the strip domain $R$ into pairs of the smaller nearly-triangular stripes in between the pole lines as follows:
\bea\label{Dsne0}
\Delta_{s} :
&&\frac{(-2\pi (s+1)+\mu\beta)\sqrt{k}}{\beta}
\le \frac{\tau_1}{\tau_2}
< \frac{(-2\pi s+\mu\beta)\sqrt{k}}{\beta} ,
\nn \\
&&\frac{(2\pi s+\mu\beta)\sqrt{k}}{\beta}
\le {\tau_1 \over \tau_2}
< \frac{(2\pi (s+1)+\mu\beta)\sqrt{k}}{\beta},
\quad s=0,1, \ldots
\eea
where poles sit on the boundaries of each $\Delta_{s}$.
The integration can be decomposed as
\bea
\int_R d^2\tau \cdots=\sum_{s=0}^\infty \int_{\Delta_s}d^2\tau \cdots
\eea
and the partition function is also decomposed accordingly
\bea
Z_1(\beta)
=\sum_{s=0}^\infty z_s(\beta) .
\eea
The integrand in each domain takes different asymptotic form.
This integral is divergent due to the poles in the integrand, so we should regularize it to apply a well-defined saddle point approximation to extract the Hagedorn behavior.
To do so, we formally remove the neighborhood of the poles of the integrand in the $\tau_2$ integral.
The explicit form of the regularized integral is in piecewise form as follows:
\be
\int_{0}^{\infty} d\tau_2 \cdots=
\lim_{\epsilon \to 0} \sum_{w=0}^\infty
\int^{\frac{\beta}{2\pi \sqrt{k}}\frac{1}{(w+\epsilon)}}_{\frac{\beta}{2\pi\sqrt{k}}\frac{1}{(w+1-\epsilon)}} d\tau_2 \cdots.
\ee
Later on we will see that these divergences from poles corresponds to the infinite warped factor felt by the long strings at spatial infinity, thus is the IR divergence. Moreover, we should point out that the form of the IR-regulator is not unique and here we choose the form for our own convenience. For example, another choice is $\int^{\frac{\beta}{2\pi w\sqrt{k}}-\epsilon}_{\frac{\beta}{2\pi(w+1)\sqrt{k}}+\epsilon}d\tau_2 \cdots$. However, it is easy to convince oneself that the different choice of the IR-regulator will not affect the leading contribution to the partition function,  
which is proportional to $\ln\epsilon$.

Note the above prescription holds for any $\mu$. However, in the following we will set $\mu$ to zero for simplicity, and then evaluate the partition function for each domain $\Delta_{s}$.
Especially we will try to extract its behavior near the Hagedorn regime.
We evaluate the contribution to the partition function from the sector $\Delta_0$ which turn out to dominate over the one from $\Delta_{s\neq 0}$  near the Hagedorn temperature.
In Appendix we argue that the contribution from $\Delta_{s\neq 0}$ is sub-dominant.

\subsubsection{Partition function from $s=0$ sector}

  Now we will extract the Hagedorn behavior from the sector $\Delta_0$ which covers the  $\tau_2$-axis and a series of poles on it due to the sin factor in $\vartheta_1$,
i.e. $\tau_1=0, \tau_2=\frac{\beta}{2\pi w\sqrt{k}}$ with $w=0, 1, \cdots$.

The contribution to the partition function from sector $\Delta_0$ is $z_0$, near the origin $\tau_1, \tau_2\simeq 0$ we can approximate $z_0$ by using the modular property
\be
|\vartheta_1(\tau,U)|\simeq 2|\tau|^{-1/2}
\Big|e^{-(U^2+1/4)\pi \tau_2/|\tau|^2} \sin{\left(\pi U \bar{\tau}\over |\tau|^2\right)}\Big|
\ee
and
\be
|\eta(\tau)|\simeq |\tau|^{-1/2} e^{-\pi \tau_2 \over 12 |\tau|^2},
\ee
where we have used the fact that the factor $1-e^{-2\pi i r/\tau}$ goes to $1$ as long as
\be
{|\tau_1| \over \tau_2} = \mbox{fixed}<{2\pi \sqrt{k} \over \beta}, \qquad \mbox{as} \quad \tau_2 \to 0.
\ee
We will see that the saddle point is located at $\tau_1/\tau_2=0$ and therefore the above constraint is consistently satisfied.

 The asymptotic behavior of the integrands ${\cal Z}_{AdS}, {\cal Z}_{int}$ and ${\cal Z}_{gh}$ are then approximated as follows:
\begin{eqnarray}
{\cal Z}_{AdS}&\simeq&
\frac{ \beta(1-2/k)^{\frac{1}{2}}}{2(4\pi^2\tau_2)^{1/2}}
 e^{-\frac{(1-2/k)\beta^2}{(4\pi\tau_2)}}\;
\frac{|\tau| e^{ (1-\frac{\beta^2}{\pi^2k})\frac{\pi\tau_2}{2|\tau|^2}}   }
{\Big|\sin \frac{i\beta\bar{\tau}}{2\sqrt{k}|\tau|^2}\Big|^2} ,
\\\label{zint1a}
{\cal Z}_{int}&\simeq&\frac{V_{int}}{(4\pi^2\tau_2)^{c_{int}/2}}
|\tau|^{c_{int}}  e^{\frac{c_{int}\pi\tau_2}{6|\tau|^2}} ,
\\
{\cal Z}_{gh}&\simeq&|\tau|^{-2}e^{\frac{-2\pi\tau_2}{6|\tau|^2}} .
\end{eqnarray}
Note that the leading exponent in \eq{zint1a} is universal for compact unitary CFTs by the property of the modular invariance of the partition function \cite{Cardy}.  Moreover, it is easy to verify for oneself from the calculations below that the Hagedorn temperature will only depend on this exponent and will be universal \footnote{We thank the authors in \cite{Berkooz:2007fe} for pointing this out.}. On the other hand, the other Hagedorn thermodynamic quantities will depend on also the non-universal part of \eq{zint1a}.  

Under the above approximation, the partition function becomes
\bea\label{z0app}
z_0(\beta)
\simeq
\frac{V_{int}}{2}
\lim_{\epsilon \to 0} \sum_{w=0}^\infty
\int^{\frac{\beta}{2\pi \sqrt{k}}\frac{1}{(w+\epsilon)}}_{\frac{\beta}{2\pi\sqrt{k}}\frac{1}{(w+1-\epsilon)}}
\frac{d\tau_2}{4\tau_2}
\frac{\beta(1-2/k)^{1/2}}{(4\pi^2\tau_2)^{(c_{int}+1)/2}}
\exp\left[-\frac{(1-2/k)\beta^2}{4\pi\tau_2}\right]
I(\tau_2)
\eea
where
\bea\label{I1eq}
I(\tau_2):=
\int_{-1/2}^{1/2} d\tau_1|\tau|^{c_{int}-1}
\Big|\sin\left(\frac{i\beta\bar{\tau}}{2\sqrt{k}|\tau|^2}\right)\Big|^{-2}
\exp\left[\frac{(c_{int}+1-3\beta^2/k\pi^2)}{6}\frac{\pi\tau_2}{|\tau|^2}\right].
\eea
Then we will evaluate the $\tau_1$ integral with fixed $\tau_2$ by the saddle-point approximation.
 Let us introduce
$x:=\tau_1/ \tau_2$,
and we can write
\bea\label{Isaddle}
I(\tau_2)
=
\tau_2^{c_{int}}\int_{-1/2\tau_2}^{1/2\tau_2} dx
\exp(-G(x;\tau_2))
\eea
where
\bea
G(x;\tau_2)=-\frac{c_{int}-1}{2}\ln(1+x^2)
+2\ln\Big|\sin\left(\frac{\beta(1+ix)}{2\sqrt{k}\tau_2(1+x^2)}\right)\Big|
-\frac{\pi(c_{int}+1-3\beta^2/k\pi^2)}{6\tau_2(1+x^2)} .
\eea
It is easy to see that
\bea
G'(x;\tau_2)&=&\frac{(1-c_{int})x}{1+x^2}
+\frac{\pi(c_{int}+1-3\beta^2/k\pi^2)x}{3\tau_2(1+x^2)^2}
\nonumber \\\label{F1}
&&
+\cot\left(\frac{\beta(1+ix)}{2\sqrt{k}\tau_2(1+x^2)}\right)\frac{\beta}{2\sqrt{k}\tau_2}
\left[\frac{-2x(1+ix)}{(1+x^2)^2}+\frac{i}{1+x^2}\right]+c.c.
\eea
Though the expression \eq{F1} looks rather complicated,
it is straightforward to find that $x=0$ is an extremal point such that  $G'(0;\tau_2)=0$.
Thus we have
\bea\label{I1f}
I(\tau_2)
\simeq
\tau_2^{c_{int}}
\Big|\sin\left(\frac{\beta}{2\sqrt{k}\tau_2}\right)\Big|^{-2}
\exp\left[\frac{\pi(c_{int}+1-3\beta^2/k\pi^2)}{6\tau_2}\right]
\sqrt{\frac{2\pi}{G''(0;\tau_2)}} .
\eea
Note that as in the case of flat space the $1/\tau_2$ factor in the exponential of \eq{I1f} provides the required form for the Hagedorn partition function when combining with the similar factor in \eq{z0app}, but the additional dressing $\sin$ factor encodes the pole structure of the partition function rendering the discrete Hagedorn spectrum, which is different from the case in flat space.

Furthermore, to ensure that $x=0$ is a stable saddle point we should require
\bea
G''(0;\tau_2)=1-c_{int}
+\frac{\pi}{3\tau_2}(c_{int}+1-\frac{3\beta^2}{k\pi^2})
-\frac{\beta}{2\sqrt{k}\tau_2}
\left[4\cot\left(\frac{\beta}{2\sqrt{k}\tau_2}\right)
-\frac{\beta/\sqrt{k}\tau_2}{\sin^2(\beta/2\sqrt{k}\tau_2)}\right]>0.
\label{G''}
\eea
Since $G''(0;\tau_2)$ is positive only for small $\tau_2$, we need to impose a lower bound $w_{min}$ in the $w$-summation of \eq{z0app} or equivalently an upper bound $\tau_2^{max}$ in the $\tau_2$ integration, so that
\be\label{G''0}
G''(0,\tau^{max}_2)=0
\ee
then
\be
w^{min}(\beta):=1+\left[{\beta \over 2\pi \tau^{max}_2 \sqrt{k}}\right]
\ee
where $\left[\cdots\right]$ is the Gauss's symbol. This then excludes the $w=0$ mode which locates at $\tau_2=\infty$, thus $w_{min}\ge 1$.

 Numerically we find that there are two more unstable extremal points other than $x=0$.
However, in the $\tau_2 \to 0$ limit, i.e. the Hagedorn regime, they are sub-dominant in the partition function as compared with the contribution from $x=0$.
This is similar to the flat space string in which the subdominant saddle points give only power law of $\tau_2$ for the corresponding $I(\tau)$, thus it is not compatible with the exponential blow-up of the Hagedorn spectrum.

The partition function can then be put into a form which highlights the Hagedorn behavior as follows
\bea
z_0(\beta)
&\simeq&
\frac{V_{int}\beta(1-2/k)^{1/2}}{8(4\pi^2)^{(c_{int}+1)/2}}
\lim_{\epsilon \to 0} \sum_{w=w_{min}}^{\infty}
\int^{\frac{\beta}{2\pi \sqrt{k}}\frac{1}{(w+\epsilon)}}_{\frac{\beta}{2\pi\sqrt{k}}\frac{1}{(w+1-\epsilon)}}d\tau_2
\tau_2^{(c_{int}-3)/2}
\sqrt{\frac{2\pi}{G''(0;\tau_2)}}
\nonumber \\\label{z0H}
&&\times
\Big|\sin\left(\frac{\beta}{2\sqrt{k}\tau_2}\right)\Big|^{-2}
\exp\left[-\frac{\beta^2-\beta_{AdS}^2}{4\pi\tau_2}\right] ,
\eea
where
\bea\label{maint1}
\beta_{AdS}:= \sqrt{\frac{2\pi^2 (c_{int}+1)}{3}}
=4\pi \sqrt{\frac{k-9/4}{k-2}}.
\eea
Recall that we have set $\alpha'=1$. The exponent factor in \eq{z0H} takes the standard form of the Hagedorn spectrum, namely, it is suppressed as $\tau_2 \to 0$ if $\beta > \beta_{AdS}$, we can then identify $\beta_{AdS}$ as the inverse Hagedorn temperature of string theory in AdS$_3 \times \cM$ \footnote{For similar arguments on the effect of background fields on the Hagedorn temperature, see e.g. \cite{Ferrer:1990na} \cite{Russo:1994cv}.}. This Hagedorn temperature is universal for compact internal unitary CFTs.

 The result shows that the Hagedorn temperature is monotonically decreasing as $k$ grows.
In the large $k$ limit this becomes the Hagedorn temperature in the flat space, as it should be. In the small $k$ regime the $\alp$ effect on the background is important since the AdS curvature scale is large.
We find that there is a lower bound for $k$, i.e. $k\ge 9/4$ in order to have Hagedorn behavior occurring at finite critical temperature $1/\beta_{AdS}$, otherwise $\beta^2_{AdS}$ is negative. It is interesting to see that $k_0\approx 2.26$ in \eq{k0e} for unitary internal CFT is slightly larger than $9/4$. 
This means that the Hagedorn temperature cannot be infinite if the internal CFT is unitary. The maximal Hagedorn temperature is about $0.388$ $l_s^{-1}$ at $k=k_0$.  

   In summary, though the inverse Hagedorn temperature in AdS$_3$ depends on $k$, it is of order of the string scale for all physical values of $k$ consistent with the unitarity constraint of the internal CFT.

Furthermore we can approximate (\ref{G''}) in the region where $\tau_2$ is very small, and the result is
\footnote{
However, this is not the leading contribution when we consider the flat limit $k \to \infty$.
The leading behavior in the limit is given by
\bea
G''(0;\tau_2)\simeq\frac{\pi}{3\tau_2}\left(c_{int}+1-\frac{3\beta^2}{k\pi^2}\right) .
\eea
}
\bea
G''(0;\tau_2) \simeq 2\left[\frac{\beta/2\sqrt{k}\tau_2}{\sin(\beta/2\sqrt{k}\tau_2)}\right]^2 .
\eea
The partition function then becomes
\bea
z_0(\beta)
\simeq
\frac{\sqrt{\pi}V_{int}(k-2)^{1/2}}{4(4\pi^2)^{(c_{int}+1)/2}}
\lim_{\epsilon \to 0} \sum_{w=w_{min}}^{\infty}
\int^{\frac{\beta}{2\pi \sqrt{k}}\frac{1}{(w+\epsilon)}}_{\frac{\beta}{2\pi\sqrt{k}}\frac{1}{(w+1-\epsilon)}}d\tau_2
\tau_2^{(c_{int}-1)/2}
\frac{\exp\left[-\frac{\beta^2-\beta_{AdS}^2}{4\pi\tau_2}\right]}
{\Big|\sin\left(\frac{\beta}{2\sqrt{k}\tau_2}\right)\Big|} .
\eea
It is convenient to change the variable by introducing $y:=\beta/2\pi\sqrt{k}\tau_2$ so that the poles due to the $\sin$ factor locate at $y=w, w=w_{min},w_{min}+1,\cdots$, which is now excluded from the integration by the cut-off.
The partition function becomes
\bea\label{z0y}
z_0\simeq
\frac{\sqrt{\pi}V_{int}(k-2)^{1/2}}{4(4\pi^2)^{(c_{int}+1)/2}}
\lim_{\epsilon \to 0} \sum_{w=w_{min}}^\infty \int^{w+1-\epsilon}_{w+\epsilon}
\frac{dy}{y}\left(\frac{\beta}{2\pi\sqrt{k}y}\right)^{(c_{int}+1)/2}
\frac{\exp\left[-\frac{\beta^2-\beta_{AdS}^2}{2\beta}\sqrt{k}y\right]}
{|\sin\left(\pi y\right)|}.
\eea
 From this, we can formally read off the (single-string) density of state of the Hagedorn spectrum for the thermal AdS$_3$ string when $\beta\sim \beta_{AdS}$ as follows
\be
\rho_{AdS_3}(E) \sim E^{-{c_{int}+3 \over 2}}
\Big|\sin\left({\pi E\over \sqrt{k}}\right)\Big|^{-1} e^{\beta_{AdS}E}.
\ee
Compare with the flat space Hagedorn density of states, 
there is an additional $\sin$ factor in $\rho_{AdS_3}$ which renders infinite number of poles in it. From the spectrum analysis done in \cite{Maldacena:2000kv}, we know that these poles are related to the long string configurations which are absent in flat space. The divergences associated with these poles are nothing but the infinite AdS$_3$ warped factor at spatial infinity felt by the space-like long strings, which corresponds to the infinite AdS$_3$ volume.  Note that the IR divergence of the flat space string is explicitly represented by the zero modes in the path-integral. On the contrary, in AdS space the zero modes are not explicit in the WZW model formulation due to the lack of the translational invariance in the radial direction, and we need to extract the IR divergence corresponding to the infinite AdS volume by manipulating the integration over poles in \eq{z0y}.

It is obvious that the main contribution in each integration in (\ref{z0y}) is given at the end points of the integration and it behaves as $\ln \epsilon$ which relates to the aforementioned IR divergence in the partition function.
To see this explicitly, we expand the $\sin$ factor around $y=w$ with new integration variable $t=y-w$ :
\bea
|\sin\left(\pi (w+t)\right)|=|(-1)^w\pi t+ \cdots|.
\eea
Therefore we can extract the divergence of the partition function as
\bea
\lim_{\epsilon\rightarrow 0}\left(
\int^{-\epsilon}\frac{dt}{| (-1)^w\pi t|}+\int_{\epsilon}\frac{dt}{|(-1)^w\pi t|}
\right)Y(w+t)
\sim \lim_{\epsilon\rightarrow 0} \left(\frac{2}{\pi}\ln \epsilon\right)Y(w) ,
\eea
where
\bea
Y(y)=
\frac{\sqrt{\pi}V_{int}(k-2)^{1/2}}{4(4\pi^2)^{(c_{int}+1)/2}y}
\left(\frac{\beta}{2\pi\sqrt{k}y}\right)^{(c_{int}+1)/2}
\exp\left[-\frac{\beta^2-\beta_{AdS}^2}{2\beta}\sqrt{k}y\right]
\eea
which has no singularity at $y=w$.
In this way, we can extract the IR divergence in the partition function as
\bea\label{Kbeta1}
z_0(\beta)=\frac{|\ln \epsilon|V_{int}(k-2)^{1/2}}{2\sqrt{\pi}} \left(\frac{\beta}{8\pi^3\sqrt{k}}\right)^{\frac{c_{int}+1}{2}}
\sum_{w=w_{min}}^\infty w^{-\frac{c_{int}+3}{2}}
\exp\left[-\frac{\beta^2-\beta_{AdS}^2}{2\beta}\sqrt{k}w\right] + O(\epsilon^0).
\eea

We now have a discrete spectrum after extracting the IR divergence.
In fact, the discrete spectrum can be interpreted as the long strings as follows. Near the Hagedorn temperature, we have
\be
\delta \beta^2:=\beta^2-\beta^2_{AdS}\simeq  2\beta_{AdS}(\beta - \beta_{AdS}).
\ee
We can introduce a parameter $p$ so that we can rewrite \eq{Kbeta1} in to the following form by neglecting the overall factor
\be
z_0(\beta) \sim \int_{-\infty}^{\infty} dp \sum_w  w^{-(c_{int}+4)/2} e^{-(\beta-\beta_{AdS})E}
\ee
where
\be\label{longse}
E=\left(w+ {p^2\over w k}\right)\sqrt{k}.
\ee
This is nothing but a piece of the long strings' spectrum considered in \cite{Maldacena:2000hw,Maldacena:2000kv} and $p$ can be interpreted as the momentum
along the radial direction for the long strings \footnote{The additional $\sqrt{k}$ factor in \eq{longse} is due to our different normalization in \eq {coor-t} from the one in \cite{Maldacena:2000hw,Maldacena:2000kv}. }.
This supports our interpretation of the IR divergence as the AdS volume felt by the long string at spatial infinity due to the fact that they can be at any radial position.

On the other hand, as $\beta\sim \beta_{AdS}$ we can also convert the discrete sum into a continuum one by introducing the scaling parameter $u:=w \delta\beta^2$ so that
\bea
\sum_{w=w_{min}}^\infty
=\frac{1}{\delta\beta^2}\sum_{w=w_{min}}^\infty \delta\beta^2
=\frac{1}{\delta\beta^2}\int_{\delta\beta^2w_{min}}^\infty du .
\eea
In this way, we can carry out the sum/integral to get
\bea
 z_0(\beta)=\frac{|\ln \epsilon|V_{int}(k-2)^{1/2}}{2\sqrt{\pi}} \left(\frac{\beta_{AdS}(\beta - \beta_{AdS})}{8\pi^3}\right)^{\frac{c_{int}+1}{2}}
\Gamma\left(-\frac{c_{int}+1}{2},(\beta - \beta_{AdS})\sqrt{k}w_{min}(\beta)\right),
\eea
where  $\Gamma(a,x):=\int_x^{\infty}t^{a-1} e^{-t} dt$ is the incomplete gamma function.

Though $w_{min}$ is a function of $\beta$ which may complicates thermodynamics, at Hagedorn regime one has
\be
w_{win}(\beta)\simeq w_{min}(\beta_{AdS})
\ee
which is nothing but a number and will not affect the Hagedorn critical behavior. Moreover, as $\beta \sim \beta_{AdS}$ the $w_{min}$-determining equation \eq{G''0} 
can be solved for $\tau_2^{max}$ and $w_{min}$. It is easy to verify that $w_{min}=1$ for $k_0<k\le \infty$ for which there exists the Hagedorn divergence.

\section{Thermodynamics of Hagedorn AdS$_3$ strings}
 Summarizing the results in the previous sections, the main contribution to the partition function in the Hagedorn regime is from the $s=0$ sector, which encodes the long string spectrum in the density of states. The asymptotic behavior of the resulting single-string partition function is
\be\label{z0ads}
z_0(\beta)=|\ln \epsilon|  \frac{V_{int}(k-2)^{1/2}}{2\sqrt{\pi}}\left(\frac{\beta_{AdS}(\beta - \beta_{AdS})}{8\pi^3}\right)^{\frac{c_{int}+1}{2}}
\Gamma\left(-\frac{c_{int}+1}{2},(\beta - \beta_{AdS})\sqrt{k}w_{min}\right) .
\ee
Note that $w_{min}=1$ as just mentioned.

 If we assume the single string dominance and Bose statistics of the string gas, the free energy $F_{AdS}(\beta)$ of the multi-string gas is nothing but
\be\label{fads}
F_{AdS}= -{1\over \beta} \ln Z\simeq -{1\over \beta}z_0
\ee
where $Z$ is the multi-string partition function given in \eq{multistringp}. From this, we can extract the thermodynamics of the Hagedorn string gas in AdS$_3$.

Before we discuss the thermodynamic quantities of AdS strings from the free energy, it is interesting to compare the form of \eq{z0ads} and \eq{fads} with the one for thermal bosonic string on $S\times R^d$ \cite{deotan}, namely,
\be\label{Fflat}
\beta F_{flat}=-C V_{int}(\beta-\beta_H)^{d/2}\Gamma(-{d\over 2},(\beta-\beta_H)m_0)
\ee
where $C$ is some constant and $d$ is the number of non-compact dimensions, i.e. neglecting the winding modes. We see that $F_{AdS}$ and $F_{flat}$ have the same critical behavior and singular structure except that the infrared cutoff $m_0$ in the flat space is replaced by  $\sqrt{k} w_{min}$. Therefore, the parameter $w_{min}=1$ plays the role of the infrared cutoff in the AdS space, which cannot be taken to zero.

    Based on the above similarity, we can separate the singular and regular parts of the free energy and extract the critical behavior of the free energy as in the flat space string \cite{deotan,Abel:1999rq}, and the result is
\be\label{Fadsf}
\beta F_{AdS} \simeq -h(\beta)(\beta-\beta_{AdS})^{(c_{int}+1)/2}+ \mbox{regular part}
\ee
where
\be
h(\beta)=(-1)^{c_{int}/2+1}C V_{int} |\ln \epsilon| \Gamma({c_{int}+1 \over 2}+1)^{-1} \ln [(\beta-\beta_{AdS})\sqrt{k}]
\ee
for $(c_{int}+1)/2$ is integer, and
\be
h(\beta)=(-1)^{c_{int}/2}C V_{int} |\ln \epsilon|\Gamma({c_{int}+1 \over 2}+1)^{-1}
\ee
for $(c_{int}+1)/2$ is non-integer.

From \eq{Fadsf} we can derive the other thermodynamic quantities in either canonical ensemble or micro-canonical ensemble, for the later we need to evaluate the density of states by the inverse Laplace transform of the multi-string partition function, i.e.,
\be\label{laplacet}
\Omega(E)=\int^{L+i\infty}_{L-i\infty} {d\beta \over 2\pi i} e^{\beta E} Z(\beta)
\ee
where the contour (denoted by $L$) is chosen to be to the right of all singularities of $Z(\beta)$ in the complex $\beta$ plane.  Following \cite{Brandenberger:1988aj,deotan,Abel:1999rq}, the resulting density of state for $c_{int}+1\geq 3$ is
\be\label{dos1}
\Omega(E)=CV {e^{\beta_{AdS} E+ \gamma_0 V} \over E^{(c_{int}+1)/2+1}}
\left(1+\cO(1/E)\right)
\ee
and the entropy is
\be
S=\ln \Omega(E)=\beta_{AdS} E+ n_H V + \mbox{sub-leading terms}
\ee
where $V$ is the volume of the $c_{int}+1$ ``non-compact" dimensions, 
and $\gamma_0$ and $n_H$ are constant of order $l_s^{c_{int}+1}$.  
For $0\le c_{int}+1<3$, the density of states is more complicated and the details can be found in \cite{deotan,Abel:1999rq}. 
Moreover, the specific heat evaluated from \eq{dos1} is negative and divergent at Hegedorn temperature. 
This implies that the Hagedorn thermodynamics is not well-defined, 
and one should compactify the $c_{int}-$dimensional internal space because the compactification will introduce the sub-leading Hagedorn singularities 
then make the thermodynamic ensembles well-defined.  For more discussions on the related issues, see e.g. \cite{Brandenberger:1988aj,deotan,Abel:1999rq, Nayeri}, and \cite{Salomonson:1985eq} provides an example for explicit calculation.

Another interesting point about the Hagedorn thermodynamics is how it encodes the topological information of the underlying spacetime \cite{deotan,Abel:1999rq}.
More precisely, the number of non-compact dimensions of the underlying spacetime (in the sense of neglecting the corresponding winding modes) is encoded in the critical exponent of the free energy as shown in \eq{Fflat} and \eq{Fadsf}.
In the AdS case, we see that the number of ``non-compact dimensions" encoded in \eq{Fadsf} is $c_{int}+1$ in analogue to number $d$ appearing in \eq{Fflat} for the flat space string.
The appearance of $c_{int}$ is because we have assumed the internal space to be non-compact with the translational zero-modes encoded in the exponent of the corresponding partition function \eq{zint0}.
Then one will wonder what does the extra one correspond to? We think it should correspond to the radial direction of the AdS$_3$ and the reason is as follows: Recall that in the AdS partition function \eq{AdS3Z}, 
there is no zero-mode associated with the radial and angular directions as can be read from the exponent of $\tau_2$ in \eq{AdS3Z}. However,  we see that in the Hagedorn regime the long string degrees of freedoms appear in the spectrum \eq{longse} extracted from the Hagedorn partition function \eq{Kbeta1}, where the parameter $p$ in \eq{longse} is the zero-mode associated with the momentum along the radial direction. 
The appearance of the finite energy space-like long strings \eq{longse} is due to the balance of the string tension and NS-NS B field near the AdS boundary \cite{Maldacena:2000hw}, this makes the radial direction effectively non-compact for the Hagedorn long strings.

\section{Strings/Black Hole Correspondence Principle in AdS}
\label{correspondence}
In formulating the strings/black hole correspondence in flat space, we expect that the correspondence point happens as the size of black hole is of order of the string scale \cite{Susskind:1993ws,Horowitz:1996nw}, and the Hagedorn entropy of the free thermal strings and the Bekenstein-Hawking entropy of black hole match up to a factor of order unity.
However, without taking into account of the string interaction, highly exited thus widely spreading string configurations dominate at Hagedorn regime so that it is incompatible with the Hoop conjecture of black hole formation, thus one should take into account of shrinking of the string due to the self-interactions \cite{Horowitz:1997jc} \cite{Damour:1999aw}.
We may think that the Jeans instability of the Hagedorn strings \cite{Atick:1988si} will induce a formation of black hole, with its Hawking temperature equals to the Hagedorn temperature so that the entropies match.
However we regard whether the entropy of self interacting string remains the same as that of the free string is an open problem.
In \cite{Kutasov:2005rr}, an alternative interpretation without invoking the self-interaction has been proposed as follows: the thermal string is considered to be the stretched horizon of the black hole after tachyon condensation
\footnote{
Before the condensation the local temperature on the region of the stretched horizon is above the Hagedorn temperature and the thermal tachyon develops.}. 
At the correspondence point where the Hawking temperature and the Hagedorn one match, the size of the stretched horizon spreads almost all over the space other than a small region around the vicinity of black hole which has a small size comparable to the string scale. 
Therefore, the black hole is indistinguishable from the gas of strings and the entropy of thermal strings explains the black hole entropy.

Following the discussions in flat space, it is tempting to formulate a strings/black hole correspondence in AdS space by disregarding the dynamical issues discussed in \cite{Horowitz:1997jc} \cite{Damour:1999aw}.
We can then assume that the strings/black hole correspondence in AdS space happens when the Hagedorn temperature is of order of the Hawking temperature.
It will be fulfilled automatically that the Hagedorn entropy becomes of order of the Bekenstein-Hawking entropy if the above condition holds, as can be seen from our discussions in the Introduction.  
Since the stable AdS black hole has positive specific heat, that is, the larger black hole has higher Hakwing temperature. 
Then, the correspondence principle implies that the Hagedorn strings in AdS will condense into the black hole with the size far larger than the string scale and keeping their entropy. 
This is quite different from the flat space case. 

On the other hand, we have obtained the Hagedorn temperature of strings in AdS$_3$, which has a nontrivial dependence on the AdS curvature scale $k$. 
This $k$-dependence is new to the flat space Hagedorn temperature. 
It is then interesting to see the implication of this $k$-dependence to the above conjectured correspondence principle.

First of all, we briefly recall the AdS Schwarzschild black hole in $d+1$ dimensions ($d \ge 2$) \cite{Hawking:1982dh} \cite{Witten:1998zw}.
The metric of the black hole for $d>2$ is given as
\bea
ds^2=f(r)d^2\tau +f^{-1}(r)d^2r +r^2 d\Omega^2
\label{AdSmetric}
\eea
where $f(r)=1+r^2/l_{AdS}^2-w_{d+1}M/r^{d-2}$ with $w_{d+1}=16\pi G_{d+1}/(d-1)\Omega_{d-1}$ and $G_{d+1}$ is the Newton constant in $d+1$ dimensions and $\Omega_{d-1}$ is the volume of unit $(d-1)$-sphere.
One may immediately read off its inverse Hawking temperature as a function of the Schwarzschild radius $r_+$ which is the largest solution of $f(r)=0$ :
\bea\label{betaBH2}
\beta_{BH}(r_+)=\frac{4\pi r_+}{d-2}\frac{1}{1+\frac{d}{d-2}\frac{r_+^2}{l_{AdS}^2}} .
\label{temp-radius}
\eea
\begin{figure}[h]
\begin{center}
\resizebox{12cm}{6cm}{\includegraphics{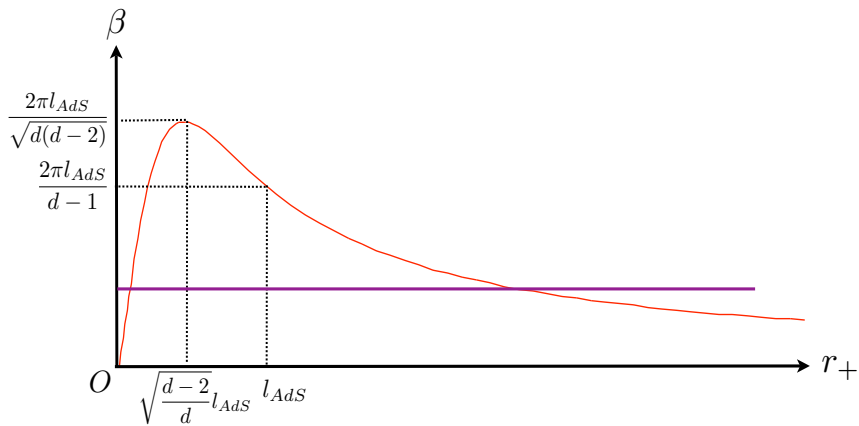}}
\end{center}
\caption{The black hole inverse temperature as a function of Schwarzschild radius in $d \geq 2$. For $d>2$ there are two solutions which correspond to the small and the large black hole.
In $d=2$, there is no solution corresponding to the small black hole. There is only a large (BTZ) black hole solution.}
\label{fig:AdSBH}
\end{figure}
We plot this function in Fig \ref{fig:AdSBH}.
Obviously, the black hole solutions exist only above the critical temperature $\beta_c^{-1}=\sqrt{d(d-2)}/2\pi l_{AdS}$. 
Below this temperature only thermal AdS space can exist.
However, if the temperature is above the critical one there exist two black hole solutions.
One is a small black hole which would reduce to an ordinary Schwarzschild black hole in the limit $l_{AdS}\to \infty$.
It has negative specific heat and thus is unstable.
Another is a large black hole which is the characteristic one in the AdS space.
This has positive specific heat and is stable (eternal black hole).
Moreover, at the temperature $\beta_{HP}^{-1}=(d-1)/2\pi l_{AdS}$ the saddle point of large black hole becomes comparable to that of the thermal AdS space 
and above the temperature it dominates the path integral. 
This is the Hawking-Page transition from thermal AdS to Schwarzschild black hole phase. 
It should be noted that the two types of black hole solution exist only for $d>2$. 
For $d=2$, we would like to emphasize that $\beta^{-1}_c=0$, there is no small black hole saddle point, only BTZ black hole solution exists at any temperature  \cite{Banados:1992wn} \cite{Banados:1992gq}.
This may be seen easily by taking $d \to 2$ limit in Fig \ref{fig:AdSBH}.

\begin{figure}
\begin{center}
\resizebox{8cm}{7cm}{\includegraphics{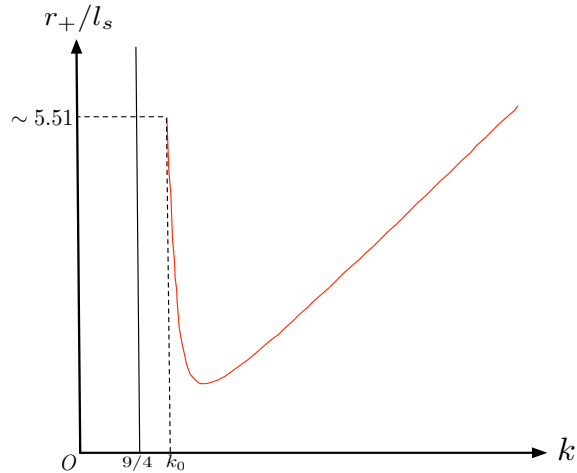}}
\end{center}
\caption{$r_+/l_s$ v.s. $k$ at the correspondence point.}
\label{fig:corresp}
\end{figure}

Let us now apply the strings/black hole correspondence to the Hagedorn strings in AdS space.
As stated earlier, if we define the correspondence point at which the Hawking and Hagedorn temperatures match, i.e., $\beta_{BH}(r_+)=\beta_{s}$, 
then the entropies of both sides coincide if string entropy is given by $\beta_{s} M$.
Accordingly, we can determine the size of the corresponding black hole at the correspondence point. 
In flat space, the correspondence principle yields a black hole of string scale. 
However, in the AdS case there are two types of black holes to which strings can correspond when we specify the string temperature. 
This can be easily seen from \eq{betaBH2} by assuming either $r_+ \ll l_{AdS}$ or $r_+ \gg l_{AdS}$. 
Upon imposing the correspondence condition $\beta_{BH}(r_{+})=\beta_{s}$, for $r_+ \ll l_{AdS}$ (small black hole with negative specific heat) we get
\bea
r_+ \simeq \frac{d-2}{4\pi}\beta_{s}\left(1+\frac{d(d-2)}{(4\pi l_{AdS})^2}\beta_{s}^2 
\right) \ll l_{AdS} ,
\label{temp-radius2}
\eea
and for $r_+ \gg l_{AdS}$ (large black hole with positive specific heat) we have
\bea
r_+ \simeq \frac{4\pi}{d}{  l_{AdS}^2\over \beta_{s}}
\eea
which implies $l_{AdS} \gg \beta_s$.  We see that in both cases, we have 
\be
l_{AdS} \gg \beta_s.
\ee
This can be seen as a prediction of the correspondence principle in AdS space in the two extreme cases we have explained above even without knowing the detailed $k$-dependence of $\beta_s$. 
Moreover, based on the semi-classical gravity the large black hole is more favored than the small one \cite{Hawking:1982dh,Witten:1998zw}, so we may expect the thermal Hagedorn strings will dynamically condensed into a large black hole.

The validity and the outcome of the above argument will depend on the explicit $k$-dependence of the Hagedorn inverse temperature $\beta_{s}$ which is unknown other than $d=2$ case obtained in this paper.
Therefore, we apply our result to the correspondence in AdS$_3$ space $(d=2)$:
Even though the metric in $d=2$ case is slightly different from (\ref{AdSmetric}), the temperature-radius relation (\ref{temp-radius}) still holds and we have
\bea\label{btzbeta}
\beta_{BTZ}=\frac{2\pi l_{AdS}^2}{r_+}=\frac{2\pi k l_s^2}{r_+}
\eea
where recall $k=(l_{AdS}/l_s)^2$. Using \eq{maint1} and \eq{btzbeta}, the correspondence condition $\beta_{BTZ}=\beta_{AdS}$ yields
\bea
\frac{r_+}{l_s}=\frac{k}{2}\sqrt{\frac{k-2}{k-9/4}} .
\eea
This relation is plotted in Fig \ref{fig:corresp}.
It is interesting to see that there is a minimum size of BTZ/string state with $r_+(k)/l_s \sim 1.69$ if  $k=(35+\sqrt{73})/16 \sim 2.72$, however, we do not have a physical understanding for such the minimum.

Obviously, for large $k$ the size of the BTZ black hole at the correspondence point is far larger than the AdS as well as the string scale.
This is in contrast to the case for the small black hole in AdS or flat space higher than the three dimensions, in which the corresponding black hole has a size comparable to the string scale.
In the limit of $k \to \infty$ the Hagedorn temperature reduces to that of the string in the flat space but the corresponding black hole has infinite size and is no longer the solution of Einstein equation. 
We only have the thermal flat space as a solution.
For the case that $k$ is infinitely large but still finite, the space becomes almost flat but still we have a large black hole phase. 
This phase might be the end point of catastrophic Hagedorn divergence through the thermal tachyon condensation in almost flat space.

On the other hand, for small $k$ close to $k_0\sim 2.26$, i.e., its minimum value for unitary internal CFT, the size of black hole grows again and reach a size of $r_+/l_s\sim 5.51$ at $k=k_0$. 
It is interesting to see that there is the ultimate black hole size at very stringy regime implied by the unitarity of the internal CFT. 
If there is no such a unitary constraint, then $k$ can reach $9/4$ which will result in a infinite size black hole.

In summary, the dependence on $k$ of our Hagedorn temperature implies that the conjectured strings/black hole correspondence in AdS$_3$ space will yield a black hole of stringy size when AdS radius is of the order of the string scale, but yield a black hole with large size when AdS radius is far larger than the string scale.

\section{Strings on BTZ}

In this section, we consider the string theory on BTZ black hole background, which is obtained from the one on thermal AdS$_{3}$ through the $SL(2, Z)$ transformation on the moduli parameter of the boundary torus \cite{Maldacena:1998bw} (see \cite{Kraus:2006wn} for a review).
According to this relation, the thermal and angular directions interchange their roles under the $SL(2, Z)$ transformation, e.g., the thermal winding modes of string in AdS$_3$ will become the angular winding modes of string in BTZ.
Thus we obtain a part of the partition function of strings on BTZ. 
Here the resulting partition function does not contain the winding modes in the thermal direction thus it may not describe whole degrees of freedom on the BTZ. However, one can expect it still captures some part of the physics on it, as we will see shortly.

\begin{figure}[h]
\begin{center}
\resizebox{8cm}{7cm}{\includegraphics{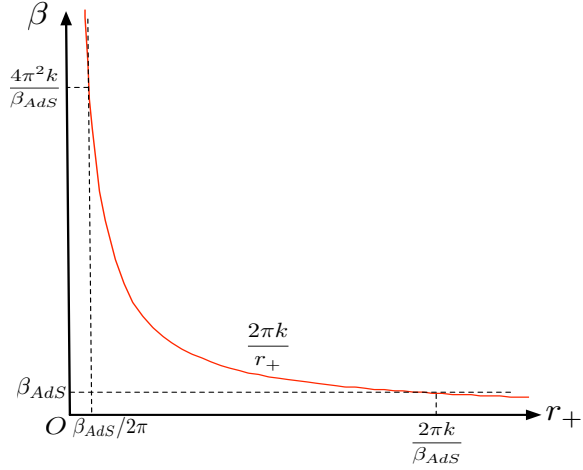}}
\end{center}
\caption{The inverse temperature-radius relation for BTZ black hole.}
\label{fig:BTZ}
\end{figure}

The moduli parameters for both sides are given
\bea
\tau_{AdS}=\frac{i \beta}{2\pi\sqrt{k}}, \quad \tau_{BTZ}=\frac{i \beta_{BTZ}}{2\pi\sqrt{k}}
\eea
respectively, and they are related through the modular transformation as $\tau=-1/\tau_{BTZ}$ then we have a relation between the inverse temperatures of strings in AdS$_3$ and on BTZ as follows
\bea\label{sl2rbeta}
\beta_{BTZ}=\frac{4\pi^2 k}{\beta} .
\eea
Thus we obtain the partition function of strings on BTZ. 
Here we denote only on the part relevant to Hagedorn divergence which is given 
\bea\label{zbtz1}
z_{BTZ}(\beta_{BTZ}) \sim 
\int d\tau_2 \cdots \exp\left(-\frac{(4\pi^2 k/\beta_{BTZ})^2-\beta_{AdS}^2}{4\pi \tau_2}\right) .
\eea
It should be noted that the exponential factor is originated from the winding modes along not the thermal but the angular direction.
This becomes diverge when
\bea
\beta_{BTZ} \geq \frac{4\pi^2 k}{\beta_{AdS}} \equiv \beta_{BTZ}^{Hag} .
\label{BTZhage}
\eea
This is the Hagedorn \footnote{We use the term Hagedorn here because the divergence in \eq{zbtz1} bears the Hagedorn form, it should be distinguished from the thermal Hagedorn divergence.}  inverse temperature of the strings on BTZ. It is interesting to see that the string ensemble goes to diverge when the temperature is lower than the $(\beta_{BTZ}^{Hag})^{-1}$ as opposed to the ordinary case.
This simply implies that BTZ black hole of low temperature is unstable due to the existence of stringy winding tachyon, even though it has a positive specific heat as implied by the gravity approximation. 
We can estimate, by noting (\ref{btzbeta}), the size of the BTZ black hole at the Hagedorn temperature as (see Fig \ref{fig:BTZ}.)
\bea
r^{(min)}_+={\beta_{AdS} \over 2\pi} .
\eea
The BTZ black hole smaller than this minimal size has a stringy halo in that the tachyon associated with the angular winding modes develops.  As noted before, $\beta_{AdS}$ is of string scale so is the $r^{(min)}_+$. This indicates the stringy $\alpha'$  correction to the gravity approximation makes the string size BTZ black hole unstable, as one will expect for string theory to be a theory of quantum gravity.

As we mentioned above our thermal partition function on BTZ is not a complete one. 
A piece of evidence for this is the partition function  \eq{zbtz1} near the Hagedorn temperature behaves as 
\be
z_{BTZ}(\beta_{BTZ}) \sim \int dE \cdots e^{(\beta_{BTZ} -\beta^{Hag}_{BTZ})E}
\ee
which has wrong Boltzmann weight, i.e. normally it should be $e^{-\beta_{BTZ} E}$,  and leads to pathological canonical ensemble entropy and micro-canonical ensemble density of states \footnote{To figure out the micro-canonical ensemble density of states, one should correctly specify the contour direction for the inverse Laplace transform \eq{laplacet}. The result is negative and thus pathological.}.
Thus though we have obtained a criterion about the minimal size of the stable BTZ black hole background based on (\ref{BTZhage}), we do not exclude a possibility that there is another Hagedorn temperature  which  gives rise to the thermal Hagedorn divergence to the BTZ partition function far above the temperature $(\beta_{BTZ}^{Hag})^{-1}$.
If this turns out to be the case, even a large black hole which is considered to be a final phase at high temperature as we argued before might not the ultimate one.

\section{Comments and Conclusion}
We would like to comment on the case with non-zero chemical potential $\mu$. One may wonder if we can generalize the above treatment for $\mu=0$ case to the $\mu\ne 0$ one. It turns out the integrand in the partition function after taking the $\tau_2 \to 0$ limit is still too complicated to perform the necessary saddle-point approximation. Here we only give the partial result for the partition function in $s=0$ sector, i.e., $z_0$.

  Similar to the $\mu=0$ case, after some analysis on the asymptotic behaviors of the factors in the theta function near the origin, we have
\bea
z_0(\beta)
&\simeq&
\frac{V_{int}}{2}
\int_{\Delta_0} \frac{d^2\tau}{4\tau_2}
\frac{\beta(1-2/k)^{1/2}}{(4\pi^2\tau_2)^{(c_{int}+1)/2}}
|\tau|^{c_{int}-1}
\Big|\sin\left(\frac{\pi U\bar{\tau}}{|\tau|^2}\right)\Big|^{-2}
\nonumber \\
&&
\times
\exp\left[\frac{(c_{int}+1)\pi\tau_2}{6|\tau|^2}
-\frac{(1-2/k)\beta^2}{4\pi\tau_2}
+\frac{\beta^2}{2\pi|\tau|^2}
\left((\mu^2-\frac{1}{k})\tau_2-\frac{2\mu \tau_1}{\sqrt{k}}\right)\right].
\eea
Unlike the $\mu=0$ case, we can not solve the saddle-point equation in closed form, therefore, it is hard to extract the Hagedorn density of states in this case. However,
the critical temperature is supposed to be read off from the exponential term by substituting
$\tau_1/\tau_2=\mu \sqrt{k}$, i.e., the ``pole line" of the $s=0$ sector, and the result is
\bea
\beta_{AdS,\mu}=\sqrt{\frac{2\pi^2(c_{int}+1)}{3(1+\mu^2 k)}}
=\frac{\beta_{AdS}}{\sqrt{1+\mu^2 k}}
\eea
where $\beta_{AdS}$ is the Hagedorn inverse temperature without chemical potential found before.
It is interesting to see that the Hagedorn temperature increases as the chemical potential increases.
We think this result is acceptable because number of state decreases if we fix angular momentum of system.
Then higher temperature than $\beta_{AdS}^{-1}$ may be required to get sufficiently large number of states
to realize the Hagedorn behavior.

 It should be emphasized that we have to justify whether $\beta_{AdS, \mu}$ is actual Hagedorn temperature in $\mu\neq 0$ case or not.
One might wonder that if other pole lines or inside region of $\Delta_s$ gives Hagedorn behavior at some temperature lower than $\beta_{AdS, \mu}$.
This is an interesting question and we leave it as an open problem.\\

In summary, we have extracted the Hagedorn behavior of thermal AdS$_3$ string from the exact 1-loop partition function. 
We find that there exists a non-trivial Hagedorn temperature given in \eq{maint1} which depends on the AdS radius scale. 
Besides, the corresponding canonical free energy and micro-canonical density of states resemble the ones for the flat space Hagedorn string with $c_{int}+1$ non-compact dimensions.
We also argue that the extra non-compact dimension encoded in the Hagedorn thermodynamics is the radial direction of AdS.
The main technical and conceptual difficulties in deriving our results are the presences of the space-like long string configurations.  
In resolving these difficulties we have carefully taken care of the IR divergence and the extract the Hagedorn behavior by a well-defined saddle point approximation.
Our results have some implication to the formulation of a correspondence principle for the formation of BTZ black hole from Hagedorn AdS strings. 
In contrast to the stringy size black hole formation from flat space Hagedorn strings,  
we find that the size of the condensed black hole could be large compared to the string scale if the AdS radius is large compared to the string scale. 
However, when the AdS radius is of order of the string scale we find that the corresponding black hole has a maximal size of order of the string scale due to the unitarity constraint for internal CFT. 
We do not know at this moment if this fact is implied by some deep truth for string theory at string scale and it deserves further study. 
We also examine strings on BTZ background obtained through $SL(2,Z)$ transformation on the boundary torus of AdS$_3$. 
We find a tachyonic divergence when a BTZ black hole is of order of the string scale and it provides a minimal size for stable BTZ black hole by taking into account the $\alpha'$ correction to the gravity approximation.

We hope our results will be helpful to understand the nature of the Hagedorn thermodynamics of string theory in the generic gravitational backgrounds.

\section*{Acknowledgements}
We are grateful to Yasuhiro Sekino for helpful discussions and inform us the work \cite{Berkooz:2007fe}. FLL also thanks Hiroshi Ooguri for inspiring short comment at earlier stage of this project during 2006 workshop at SIAS.
TM would like to thank Tsunehide Kuroki, Akitsugu Miwa, Kin-ya Oda, Yuuichirou Shibusa, Hiroshi Suzuki for valuable discussions and comments. This work is supported by Taiwan's NSC grant 94-2112-M-003-014(FLL), 95-2811-M-003-005(TM)  and 95-2811-M-003-007(DT).

\section*{Appendix :
Sub-dominant contribution form $s\ne 0$ sectors}

   Unlike the flat string case, here we have infinite number of pole lines inside the strip domain $R$. To capture the full Hagedorn thermodynamics of the AdS$_3$ string,  
we need to evaluate the contribution of these $s\ne 0$ pole lines to the partition function in each sub-strip $\Delta_{s\ne 0}$. Especially, we would like to examine the asymptotic behavior of each factor in $|\vartheta(\tau,i\beta/2\pi\sqrt{k})|^2$ given in \eq{theta12} at $\tau_2 \to 0$ near small $\tau_1$,$\tau_2$ region, 
also near the zeros of this function.

 As argued in discussing the $s=0$ sector, the factor $(1-e^{-2\pi i r/\tau})$ in $|\vartheta(\tau,i\beta/2\pi\sqrt{k})|^2$  goes to $1$ in  the small $\tau_1$, $\tau_2$ region. And in this region the sin factor behaves as
\bea
\Big|\sin\left(\frac{\pi U\bar{\tau}}{|\tau|^2}\right)\Big|
\simeq \exp\left(\frac{\beta|\tau_1|}{2\sqrt{k}|\tau|^2}\right).
\eea
The asymptotic behaviors of these two factors are independent of $s$, the remaining two factors in \eq{theta12} are not.

The exponents of the other two factors $\prod_{r=1}^\infty(1-e^{-2\pi i(r+U)/\tau})$ and $\prod_{r=1}^\infty(1-e^{-2\pi i(r-U)/\tau})$ in \eq{theta12} are
\bea\label{expne0}
\frac{1}{|\tau|^2}\left[ \tau_2\frac{\beta}{\sqrt{k}}
(-{2\pi r\sqrt{k}\over \beta} \mp  \frac{\tau_1}{\tau_2})
-i\left(2\pi r \tau_1\mp\frac{\beta \tau_2}{\sqrt{k}}\right)
\right] .
\eea
In the above, the upper-sign part corresponds to the first factor, and the lower-sign one to the second factor. Inside the sub-strip $\Delta_{s\ne0}$ defined in \eq{Dsne0}, the real part in \eq{expne0} is constrained by
\bea\label{conss0}
&& -{2\pi(s+1+r)\sqrt{k}\over \beta} < -{2\pi r\sqrt{k}\over \beta} - \frac{\tau_1}{\tau_2}\le -{2\pi (s+r)\sqrt{k}\over \beta},\\
&&{2\pi(s-r)\sqrt{k}\over \beta} \le -{2\pi r\sqrt{k}\over \beta} + \frac{\tau_1}{\tau_2}\le {2\pi (s+1-r)\sqrt{k}\over \beta}.\label{conss1}
\eea
From \eq{conss0}, we see that the real part is negative for all $r$ such that the factor $\prod_{r=0}^\infty(1-e^{-2\pi i(r+U)/\tau})$ goes to $1$ at $\tau_2 \to 0$. However, from \eq{conss1} we see that the real part is negative for $r>s$ and is positive for $r<s$, so that at $\tau_2 \to 0$ $\prod_{r=s+1}^\infty(1-e^{-2\pi i(r-U)/\tau}) \to 1$ and
\be
\Big|\prod_{r=1}^{s-1} (1-e^{-2\pi i(r-U)/\tau})\Big|\simeq \Big|\prod_{r=1}^{s-1} e^{-2\pi i(r-U)/\tau} \Big|=e^{\frac{1}{2|\tau|^2}\left(
-2\pi s(s-1)\tau_2 + 2(s-1)\frac{\beta \tau_1}{\sqrt{k}}\right)
 }.
\ee
Finally, for $r=s$ there are infinite number of poles lying on the boundary of the sub-strip, therefore we just do not further simplify it. Combined all the above, the theta function in the small $\tau_2$ region inside the sub-strip $\Delta_{s>0}$ behaves like
\bea\label{zsf}
 |\vartheta_1(\tau,U)| \simeq
|\tau|^{-1/2}
\exp\left[\frac{1}{2|\tau|^2}\left\{
\left(
\frac{\beta^2}{2\pi k}-\frac{\pi}{2}
-2\pi s(s-1) \right)\tau_2 + (2s-1)\frac{\beta \tau_1}{\sqrt{k}}\right\}
\right]
\nonumber \\
 \times
\Big|
\left(
1-\exp\left[
\frac{1}{|\tau|^2}\left[
-2\pi s \tau_2 \pm \frac{\beta \tau_1}{\sqrt{k}}
-i\left(2\pi s \tau_1\pm\frac{\beta \tau_2}{\sqrt{k}}\right)
\right]
\right]
\right)
\Big| .
\eea
Thus we have for $s \geq 1$
\bea
&& z_s(\beta)
\simeq
2V_{int}
\int_{\Delta_s} \frac{dr d\theta_s}{4\sin\theta_s}
\frac{\beta(1-2/k)^{1/2}r^{c_{int}-1}}{(4\pi^2 r \sin\theta_s)^{(c_{int}+1)/2}}
\exp\left[\frac{g(\theta_s)}{r}\right]
\nonumber \\
&&
\times
\Big|
\left(
1-\exp\left[
\frac{1}{r}\left[
-2\pi s \sin\theta_s \pm \frac{\beta \cos\theta_s}{\sqrt{k}}
-i\left(2\pi s \cos\theta_s\pm\frac{\beta \sin\theta_s}{\sqrt{k}}\right)
\right]
\right]
\right)
\Big|^{-2}
\eea
where
\bea
g(\theta_s)
=
-\frac{k-2\cos^2\theta_s}{4\pi k\sin\theta_s}\beta^2
-\frac{(2s-1)|\cos\theta_s|}{\sqrt{k}}\beta
+\frac{(c_{int}+1+12s(s-1))\pi \sin\theta_s}{6}.
\eea
 Here we have changed to the polar coordinate by defining
\be
\tau_1=r \cos \theta_s , \qquad
\tau_2=r \sin \theta_s.
\ee
In performing the integration of $\theta_s$, the contribution from the boundary would dominate the integral, thus we may regard it as a sharp saddle-point.
In fact, it would give large contributions since it contained infinitely many poles before the regularization.
The boundary of the sub-strip $\Delta_s$ is located at $\theta_s=\bar{\theta}_s$ with
\bea
\tan\bar{\theta}_s=\frac{\beta}{2\pi s \sqrt{k}}.
\eea
By setting $\theta_s$ to $\bar{\theta}_s$ in \eq{zsf} just to extract the large boundary contribution,
we have
\bea\label{zsfh}
z_s(\beta)
\simeq
\frac{V_{int}}{2}
\int_{\Delta_s} \frac{dr }{4\sin\bar{\theta}_s}
\frac{\beta(1-2/k)^{1/2}r^{c_{int}-1}}{(4\pi^2 r \sin\bar{\theta}_s)^{(c_{int}+1)/2}}
\exp\left[\frac{g(\bar{\theta}_s)}{r}\right]
\nonumber \\
\times
\Big|
\left(
1-\exp\left[
-\frac{i}{r}\left(2\pi s \cos\bar{\theta}_s\pm\frac{\beta \sin\bar{\theta}_s}{\sqrt{k}}\right)
\right]
\right)
\Big|^{-2},
\eea
where
\bea
g(\bar{\theta}_s)=
\frac{\beta(\beta_c^2-\beta^2)}{4\pi \sqrt{\beta^2+4\pi^2 s^2 k}}
\eea
with
\bea
\beta_c^2=
4\pi^2 \left(
\frac{4k-9}{k-2}
-ks^2
\right) .
\eea
If we interpret $1/r$ as the energy, then we can extract the density of states from \eq{zsfh} and examine the possible Hagedorn behavior. Note that
the second line in \eq{zsfh} contains the regularized poles but shows no exponential behavior for Hagedorn density of states. These poles can be understood as the long sting configurations discussed in \cite{Maldacena:2000hw,Maldacena:2000kv}. Instead, the Hagedorn behavior is encoded in the factor $\exp\left[\frac{g(\bar{\theta}_s)}{r}\right]$. However, $\beta^2_c$ is always negative or zero as can be easily seen. In fact, it becomes zero only when $k=3, s=1$.
Therefore, we conclude that there is no Hagedorn behavior in the sectors $\Delta_{s\ne 0}$. That is, the $\Delta_{s=0}$ sector dominates the Hagedorn behavior over the others, and we need to only consider this sector when discussing the Hagedorn thermodynamics.


 \end{document}